\documentclass[twocolumn,amsmath,amssymb,english,prb]{revtex4}

\usepackage{graphicx}% Include figure files
\usepackage{dcolumn}% Align table columns on decimal point
\usepackage{bm}% bold math
\usepackage{babel}

\begin{document}

%\preprint{APS/123-QED}

%\subjclass[pacs]{61.10.Eq, 61.10.Ht, 61.10.Nz, 61.46.+w, 68.65.Cd, 68.65.Hb}
\title{\textit{In situ} resonant x-ray study of vertical correlation and capping effects during GaN/AlN quantum dots growth}

\author{J. Coraux$^{\textrm{1,2,*}}$, H. Renevier$^{\textrm{1,2}}$, V. Favre-Nicolin$^{\textrm{1,2}}$, G. Renaud$^{\textrm{1}}$, B. Daudin$^{\textrm{1}}$}
\affiliation{$^{\textrm{1}}$D$\acute{e}$partement de la Recherche Fondamentale sur la
Mati$\grave{e}$re condens$\acute{e}$e, Service de Physique des Mat$\acute{e}$riaux
et Microstructures, CEA Grenoble, 17 rue des Martyrs 38054-Grenoble Cedex 9, France\\
$^{\textrm{2}}$Universit$\acute{e}$ Joseph Fourier, BP 53 F-38041 Grenoble Cedex 9, France\\
$^{\textrm{*}}$Johann.Coraux@cea.fr}

\date{\today}

\begin{abstract}
Grazing incidence anomalous x-ray scattering was used to monitor \textit{in
situ} the molecular beam epitaxy growth of GaN/AlN quantum dots (QDs). The
strain state was studied by means of grazing incidence
Multi-wavelength Anomalous Diffraction (MAD) in both the QDs and the
AlN during the progressive coverage of QDs by AlN monolayers. Vertical correlation in the position of the GaN
QDs was also studied by both grazing incidence MAD and anomalous Grazing Incidence Small Angle Scattering
(GISAXS) as a function of the number of GaN planes and of the
AlN spacer thickness. In a regime where the GaN QDs and the AlN capping are mutually strain influenced, a vertical
correlation  in the position of QDs is found with as a 
side-effect an average increase in the QDs width.

\end{abstract}

%\pacs{Valid PACS appear here}

\maketitle

%\section*{Introduction}

The control of growth processes is a major issue for the achievement
of room temperature optoelectronic properties in nitride 3D nanostructures
such as defect-free GaN/AlN quantum dots (QDs) \cite{Daudin97}. For
that purpose the morphological characteristics and structural properties, namely the strain,
size, size distribution, and density of QDs, must be controlled. X-ray
techniques have shown their powerful abilities to monitor \textit{in
situ} these parameters over large assemblies of QDs \cite{Renaud03}.
In this letter we report on a) an \textit{in situ} strain analysis of the capping effect during the progressive covering of QDs by AlN,
carried out with the Multi-wavelength Anomalous
Diffraction (MAD) technique in grazing incidence, and b) an \textit{in situ} study of the vertical correlation effect on the QDs size
during the progressive stacking-up of GaN QDs planes \cite{Tersoff96,Chamard01,Gogneau04}, carried out by both Grazing Incidence Small
Angle Scattering (GISAXS) and MAD.

%\section*{Experimental}

Samples were grown on 6H-SiC(0001) substrates in a plasma assisted
molecular beam epitaxy chamber installed on the French Collaborating Research
Group beamline BM32 at the European Synchrotron Radiation Facility (Grenoble, France).
The nitrogen flux was supplied by a radio-frequency plasma cell ; Knudsen cells provided the Ga and Al fluxes. The ultra-high vacuum
environment made possible the \textit{in situ} growth monitoring by
Reflection High Energy Electron Diffraction (RHEED) and x-ray probes. Prior to
GaN, a thin 8 monolayers (MLs) AlN
buffer was deposited. The AlN surface quality was controlled by RHEED. Wurtzite GaN QDs were synthetized on top of AlN via the modified
Stranski-Krastanow 
mode \cite{Gogneau03}, taking advantage of the in-plane 2.5\% lattice mismatch between AlN and GaN. The QDs were grown at $\sim 730^\circ C$ by depositing a GaN equivalent amount of 6 MLs. The first sample
consisted of one GaN QDs plane progressively covered by 2, 4, 6, 8, 10, 12, 20,
27, 34 MLs of AlN. The second sample was a progressive stacking of 10 GaN QDs
planes separated by 25 MLs AlN spacers. The third sample was made of 2 successive
progressive stacks of 5 QDs planes separated by a) 25 MLs and b) 53 MLs AlN spacers.

A grazing incidence and exit setup was used to enhance the scattered x-ray signals from the QDs with respect to that of the substrate.
Diffraction anomalous measurements were carried out around the Ga K-edge (10.367 keV). For the MAD
experiments reciprocal space scans along
the $[10\bar{1}0]$ direction around  \textit{h}=3 were systematically recorded at 12 energies around the Ga K-edge using a point
detector. The detector slits were opened so as to measure the integrated intensity over the grazing exit angle $\alpha_{f}$. A 2D
charge-coupled device detector placed perpendicular to the incident beam was used for GISAXS measurements \cite{Renaud03}.

%\section*{Results and Discussion}

\begin{figure}
\includegraphics[width=8cm,keepaspectratio]{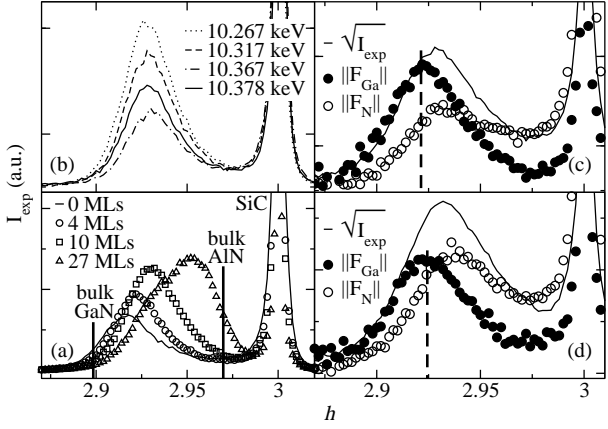}
\caption{\label{fig:AlN capping}(a) : \textit{h}-scans along $[10\bar10]$ at 10.267 keV, sub-critical
incident angle ($\alpha_{i}=0.17^\circ$),
for increasing AlN coverage. The SiC substrate peak remains at \textit{h}=3 while
the GaN QDs peak progressively shifts to higher \textit{h} values. Vertical lines indicate the bulk GaN
and AlN. (b) : Experimental $I_{exp}$ measured at 4 energies across the Ga K-edge for a 8 MLs AlN coverage. (c), (d) :
$\sqrt{I_{exp}}$
measured at 10.267 keV, $\left\Vert F_{Ga}\right\Vert $ and $\left\Vert F_{N}\right\Vert$ extracted for a 8 MLs (c) and a 12 MLs (d) AlN coverage.}
\end{figure}

Figure \ref{fig:AlN capping}(a) shows the evolution of \textit{h}-scans close to
the $\left(30\bar{3}0\right)$ reflection as a function of the AlN coverage. These \textit{h}-scans are related to both the
in-plane strain state and size. The incident x-ray angle was fixed at $0.17^\circ$, under the
critical angle for total external reflection ($\sim 0.23^\circ$ for bulk GaN in this energy range), so as to limit the scattering length
below 10 nm and have surface sensitive measurements. With no AlN coverage, one observes
a diffuse-scattering peak ascribed to QDs slightly strained by the AlN
buffer and SiC substrate. As the AlN coverage increases, this peak is progressively distorted and
shifted towards bulk AlN.

Further analysis was made possible by distinguishing the GaN contribution from the AlN and SiC contributions, using MAD measurements
\cite{Letoublon04,Hodeau01,Hendrickson91}. Figure \ref{fig:AlN capping}(b) shows some of the diffracted intensities measured \textit{in
situ} for a 8 MLs AlN coverage, at 12 energies across the Ga K-edge, taking advantage of the Ga anomalous effect to localize
the Ga contribution along $\left[10\bar10\right]$. For these measurements, the Ga partial structure factor $F_{Ga}$ of phase
$\varphi_{Ga}$, that includes the Thomson scattering of all anomalous atoms (Ga), can be retrieved. In principle, the retrieval should be run 
in the framework of the Distorted Wave Born Approximation \cite{Pietsch04,Schmidbauer05}. However, one can discard the
energy dependence at the Ga K-edge of the reflection coefficients, as a consequence of the small Ga amount (6 equivalent MLs). Therefore, the
recorded intensity corrected for fluorescence, $I_{exp}$, is proportional to the total square structure factor $\left\Vert F \right\Vert ^{2}$. 

%\begin{equation}
%I_{exp} \left(E\right) \propto \left\Vert F\right\Vert ^{2} \propto \left\Vert F_{T}\right\Vert ^{2}\times
%\left\lbrace \left[\mathrm{cos}\left(\varphi_{T}-\varphi_{Ga}\right)+
%\beta f'_{Ga}\right]^{2}+\left[\mathrm{sin}\left(\varphi_{T}-\varphi_{Ga}\right)+\beta f''_{Ga}\right]^{2}\right\rbrace
%\label{eq:Iexp}
%\end{equation}

\begin{eqnarray}
I_{exp} \left(E\right) \propto \left\Vert F\right\Vert ^{2} \propto \left\Vert F_{T}\right\Vert ^{2}\times\nonumber\\
\left\lbrace \left[\mathrm{cos}\left(\varphi_{T}-\varphi_{Ga}\right)+
\beta f'_{Ga}\right]^{2}+\left[\mathrm{sin}\left(\varphi_{T}-\varphi_{Ga}\right)+\beta f''_{Ga}\right]^{2}\right\rbrace
\label{eq:Iexp}
\end{eqnarray}

where $\beta=\left\Vert F_{Ga}\right\Vert /\left(f_{Ga}^{0}\left\Vert F_{T}\right\Vert\right)$. The partial structure factor $F_{T}$ of
phase $\varphi_{T}$ that includes the overall contribution of non anomalous atoms and the Thomson scattering of all anomalous atoms,
$F_{Ga}$, as well as $\varphi_{T}-\varphi_{Ga}$, can be extracted for all \textit{h} values, without any structural model by
fitting eq. \ref{eq:Iexp} to the experimental data with the NanoMAD algorithm \cite{FavreNicolin}.

As shown in fig. \ref{fig:AlN capping}(c,d), $F_{Ga}$ was extracted and $F_{N}$, the complex structure factor of non anomalous atoms, was
deduced from the  extracted $F_{T}$ and $\varphi_{T}-\varphi_{Ga}$, taking into account the wurtzite crystallographic structure. For
\textit{h} smaller than 2.985, \textit{i.e.} far from the SiC substrate peak, only Al and N contribute to $F_{N}$. The
\textit{h} position of the diffuse $F_{Ga}$ and $F_{N}$ peak maximum is inversely
proportional to the in-plane lattice parameter, since $\left(h-h_{SiC}\right)/h_{SiC}=\left(a-a_{SiC}\right)/a_{SiC}$ where
$a_{SiC}\simeq\,3.081 \AA$ and $h_{SiC}\simeq\,3$.

\begin{figure}
\includegraphics[width=7cm,keepaspectratio]{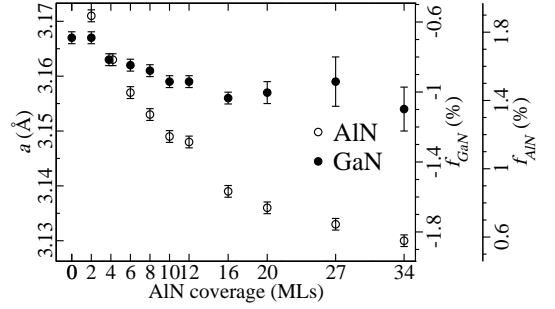}
\caption{\label{fig:aAlN aGaN}In-plane lattice parameter and mismatches (relative to bulk AlN or GaN) for AlN and GaN deduced
from $F_{N}$ and $F_{Ga}$ extractions. The error bars in the last 3 points for GaN are large due to weak extracted $F_{Ga}$.}
\end{figure}

Figure \ref{fig:aAlN aGaN} depicts the evolution of the in-plane lattice parameters and mismatches in both AlN and GaN,
as a function of the AlN cap layer thickness. The lattice mismatch in the material, relative to the
corresponding bulk material, is $f=\left(a-a_{bulk}\right)/a_{bulk}$. It is clearly demonstrated that the AlN capping initially stressed
by the QDs ($f_{AlN}\simeq\,+1.7\%$) is then progressively relaxed. Two different regimes can be distinguished, in which QDs and AlN
capping are mutually influenced. A rapid decrease is observed until 16 MLs, followed by a more gentle diminution probably
lasting after 34 MLs. Even for large AlN deposits (100 nm, not shown in Fig. \ref{fig:aAlN aGaN}) AlN remains slightly expanded
($f_{AlN}\simeq\,+0.6\%$). Jointly the partially relaxed ($f_{GaN}\simeq\,-0.7\%$) surface QDs are gradually compressed by the AlN capping
untill 16 MLs, leading to a plateau at $\sim -1.1\%$, keeping in mind the large uncertainties for AlN coverage thicker than 16 MLs.
Interestingly, the AlN strain state varies only very slowly above an AlN thickness of $\sim$ 30 MLs (7.5 nm).
Under this semi-quantitative limit, the AlN tickness is such that the presence of buried dots significantly affects the AlN strain
state, and a GaN QDs vertical correlation is expected.

\begin{figure}
\includegraphics[width=7cm,keepaspectratio]{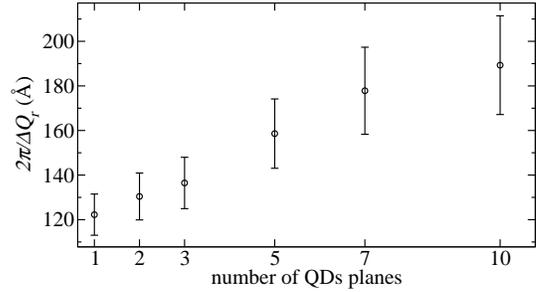}
\caption{\label{fig:DGaN}$2\pi/\Delta Q_{r}$ as a function of the number of QDs planes.}
\end{figure}

Another grazing incidence MAD study has been carried out for a gradual QDs planes stacking. The AlN spacer was chosen in a regime of
mutual QDs and AlN influence and fixed at 25 MLs. $F_{Ga}$ and $F_{N}$ were extracted for 1, 2, 3, 5, 7, and 10 capped QDs planes, in
order to obtain an information regarding the finite-size of GaN domains. Indeed, the full width at half maximum $\Delta Q_{r}$ of $F_{Ga}$ in radial
scans is related to the size
$D_{GaN}$ by the Scherrer-like formula\begin{equation}
D_{GaN}\propto\frac{2\pi}{\Delta Q_{r}}\label{eq:scherrer}\end{equation}

where the broadening along the scattering vector $\Delta Q_{r}=\Delta h/\left(\sqrt{3}a_{SiC}\right)$,
$\Delta h$ being the width of the $F_{Ga}$ peak along the \textit{h} direction.
Equation (\ref{eq:scherrer}) assumes that the broadening by a strain distribution
effect is negligible \cite{Renaud99}. The evolution of $D_{GaN}$
with the number of plane in the stack is plotted in Figure \ref{fig:DGaN}.
Despite the 10\% uncertainties, an increase of $\sim$50\%
in the size of QDs is put in evidence, as a consequence of a vertical
correlation effect.

\begin{figure}
{\includegraphics[width=8cm,keepaspectratio]{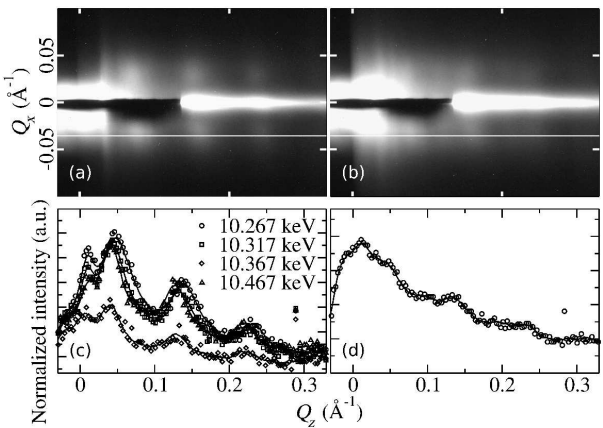}}
\caption{\label{fig:GISAXS}(a,b) : GISAXS maps measured at 10.267 keV and $\alpha_{i}=0.23^\circ$ (larger than the critical angle in
AlN) for 5 QDs planes stackings with (a) 25 MLs and (b) 53 MLs AlN spacers. $Q_{z}$ is along the growth direction $[0001]$.
 $Q_{x}$ stands for the in-plane direction. (c,d) : Experimental (circles) and smoothed (lines) GISAXS intensities for
$Q_{x}=-0.033 \AA^{-1}$ (white line in (a,c)). (d) is measured at 10.267 keV. The $Q_{z}$ period observed in (a,c) correspond to 6.8 nm,
\textit{i.e.} $\sim$ the period of the stack along $\left(0001\right)$. The faint periodic behaviour present in (b,d), which period is the same
as in (a,c), originates from the stacking with 25 MLs AlN spacers underneath.}
\end{figure}

The vertical correlation effects were further investigated by anomalous
GISAXS. The incident angle was set slightly above the critical angle,
making the technique sensitive to the whole stacking. Figure \ref{fig:GISAXS}(a)
shows GISAXS measurements for a stack of 5 QDs planes separeted by 25 AlN MLs. The occurrence of $Q_{z}$ satellites at
$Q_{x}=\,\pm0.033 \AA^{-1}$ is an evidence for the QDs position vertical correlation, \textit{i.e.} along the $\left[0001\right]$ direction
\cite{Stangl04,Chamard03}. Note that the period of these satellites, $\Delta Q_{z}$, corresponds to a stack period
$\Omega=2\pi/\Delta Q_{z}\simeq\, 6.8 nm$, \textit{i.e.} the bilayer thickness. The strong energy dependence at the Ga K-edge of these satellites (Fig. \ref{fig:GISAXS}(c))
confirms that the scattering yield is mainly from the QDs. Finally, fig. \ref{fig:GISAXS}(b) shows a GISAXS
map measured for a QDs planes stacking with 53 MLs AlN spacers. Compared to the
case of thiner (25 MLs) AlN spacers (Fig. \ref{fig:GISAXS}(a,c)) 53 MLs AlN spacers (Fig. \ref{fig:GISAXS}(b,d))
seem large enough to inhibit the vertical correlation effects, likely because they are
out of the regime of QDs and AlN mutual influence. This is consistant with the above conclusion that the critical
thickness of the AlN spacer requested to allow correlation lays around 30 MLs.

In summary, the \textit{in situ} diffraction analysis of the capping has shown that the strain state for the AlN spacer is influenced by
the buried QDs up to 30 MLs of AlN capping, which corresponds to the upper limit (7.5 nm spacers) to observe correlation between
successive GaN layers. More specifically, the connection
between capping effects and vertical correlation effects was put in evidence.
In particular it was shown that in a regime where the GaN QDs and the AlN capping are mutually strain influenced, a vertical correlation
in the position of QDs is enabled with, as a 
side-effect, an average increase in the QDs width.

The authors would like to thank Y. Genuist, Y. Cur$\acute{e}$, M. Lafossas, and M. Noblet-Ducruet for their precious technical
assistance.

%\newpage

%width=0.80\textwidth
%width=5cm,keepaspectratio

\bibliography{/mnt/win_d/these/articles/GaN_insitu_applied_physics_letters/GaN_insitu}

\end{document}